\documentclass[twocolumn,american,english,aps,prl]{revtex4}
\usepackage[T1]{fontenc}
\usepackage[latin9]{inputenc}
\usepackage{color}
\usepackage{babel}
\usepackage{textcomp}
\usepackage{bm}
\usepackage{amsmath}
\usepackage{amssymb}
\usepackage{graphicx}
\usepackage{esint}
\usepackage[unicode=true,pdfusetitle,
 bookmarks=true,bookmarksnumbered=false,bookmarksopen=false,
 breaklinks=false,pdfborder={0 0 1},backref=false,colorlinks=false]
 {hyperref}

\makeatletter
\@ifundefined{textcolor}{}
{%
 \definecolor{BLACK}{gray}{0}
 \definecolor{WHITE}{gray}{1}
 \definecolor{RED}{rgb}{1,0,0}
 \definecolor{GREEN}{rgb}{0,1,0}
 \definecolor{BLUE}{rgb}{0,0,1}
 \definecolor{CYAN}{cmyk}{1,0,0,0}
 \definecolor{MAGENTA}{cmyk}{0,1,0,0}
 \definecolor{YELLOW}{cmyk}{0,0,1,0}
}

\makeatother

\begin{document}
\title{Single-particle versus collective effects in assemblies of nanomagnets:
screening}
\author{F. Vernay and H. Kachkachi}
\affiliation{Université de Perpignan Via Domitia, Laboratoire PROMES-CNRS (UPR-8521),
Rambla de la Thermodynamique, Tecnosud, 66100 Perpignan, FRANCE.}
\date{\today}
\begin{abstract}
We discuss experimentally realizable situations in which surface effects
may ``screen out'' the dipolar interactions in an assembly of nanomagnets,
which then behaves as a noninteracting system. We consider three examples
of physical observables, equilibrium magnetization, ac susceptibility
and ferromagnetic resonance spectrum, to illustrate this screening
effect. For this purpose, we summarize the formalism that accounts
for both the intrinsic features of the nanomagnets and their collective
effects within an assembly the condition for screening.
\end{abstract}
\maketitle

\section{Introduction}

For many years research on assemblies of nanomagnets (NM) has been
devoted to the investigation of the two aspects: i) the single-nanomagnet
(SNM) physics that is mostly governed by the finite size and boundary
effects which become rather acute at the nanometer scale, ii) the
collective behavior of the assembly (ANM) that is mostly driven by
the mutual long-range dipolar interactions. Numerous studies, experimental
and theoretical, have been carried out in order to investigate the
interplay between the phenomena that occur at these two rather different
scales: While the relevant temperature at the SNM scale is about a
few degrees (Kelvin), it is about a few tens of degrees for the ANM.
While the relevant relaxation time is about a nanosecond for the former,
it is about a few minutes or hours for the latter, and so on.

From the technological perspective, many challenging issues are still
open. For example, to what extent assembling into macro-scaled systems
and devices various kinds of nano-objects may preserve their novel
properties pertaining to the nanoscale? From a fundamental viewpoint,
the foremost aim is to achieve a clear understanding of the behavior
of the nano-scaled systems and this requires the development of new
and adequate tools for their probe, measurement and modeling. Indeed,
the transition from the nano-scale to the macro-scale requires a fair
understanding of the interplay between the intrinsic properties of
the nano-objects, taken separately, and their mutual interactions
when they are assembled into more complex structures. An example of
paramount importance for practical applications is concerned with
the dynamics of the assembly and how it is related to the SNM dynamics.
This issue has been investigated for decades and the debate is still
at its climax, for instance in what regards the conditions under which
a spin-glass behavior or a superferromagnetic order may be observed,
for a given set of physical parameters of the individual NM.

Conversely, one may address the question as to whether it is possible
to define a set of such SNM parameters, on one hand, and those pertaining
to the ANM, on the other, so that somehow the ANM would behave as
a free assembly of ``dressed'' NM. In other terms, the intrinsic
properties of the individual NM, such as surface effects (SE), would
screen out the dipolar interactions (DI) within the assembly. For
example, it has been shown by many researchers \citep{strijkersetal99jap,ancinasetal01jap}
that one can control the inter-particle interactions by tuning the
properties of the NM themselves. On the other hand, it is possible
to modify the NM properties by modifying the assembly characteristics
(shape, spatial organization, etc) \citep{troetal03jmmm,troncetal00jmmm}.

In the present work, we contribute to this study by establishing the
conditions, realizable in experiments, under which the above-mentioned
screening may occur. For this, we first summarize our previous work
that has been carried out with the objective to build a global picture
that encompasses both what is happening at the SNM and at the assembly
level. Indeed, we have built a complete theoretical toolbox for dealing
with the collective behavior of an ANM whiling taking account of the
NM features.

The paper is organized as follows: in Section \ref{sec:Single-nanomagnet}
we summarize our formalism for the single SNM and the corresponding
effective model that accounts for the intrinsic properties such as
SE. In Section \ref{sec:NMA-FS} we discuss the DI in a formalism
that takes account of the finite-size of the NM and is compared to
the point-dipole approximation. In Section \ref{sec:SN-vs-NMA} we
consider in three successive paragraphs the equilibrium magnetization,
the ac susceptibility and the ferromagnetic resonance spectrum. In
each case we discuss the screening of the DI by the SE and establish,
when analytically possible, the corresponding conditions. The final
section is devoted to conclusions and discussion.

\section{\label{sec:Single-nanomagnet}Single nanomagnet}

For the single NM considered as a crystallite of $\mathcal{\ensuremath{N}}$
atomic magnetic moments $\bm{m}_{i}=m\bm{s}_{i}$ ($\left\Vert \bm{s}_{i}\right\Vert =1$),
it is possible in principle to investigate, to some extent, most of
the magnetic properties (at equilibrium and out-of-equilibrium)\citep{dimwys94prb,kodber99prb,kacgar01physa300,kacgar01epjb,igllab01prb,kacdim02prb,kacgar05springer,kazantsevaetal08prb}.
This is usually done with the help of the atomistic approach based
on the anisotropic (classical) Heisenberg Hamiltonian

\begin{equation}
\mathcal{H}=-\frac{1}{2}\sum\limits _{i,j}J_{ij}\,\mathbf{s}_{i}\cdot\mathbf{s}_{j}-\mu_{a}\mathbf{H}\cdot\sum_{i=1}^{\mathcal{N}}\mathbf{s}_{i}-\sum_{i=1}^{\mathcal{N}}K_{i}\,\mathcal{A}(\mathbf{s}_{i})\label{eq:Ham-MSP}
\end{equation}
with the usual meaning for the the microscopic parameters $J,K$.
$\mathcal{A}(\mathbf{s}_{i})$ is the anisotropy function that depends
on the locus of the atomic spin $\mathbf{s}_{i}$. So for core spins,
the anisotropy may be uniaxial and/or cubic, while for surface spins
there are various models for on-site anisotropy that is very often
taken as uniaxial with either a transverse or parallel easy axis.
There is also the more plausible model of Néel for which $\mathcal{A}(\mathbf{s}_{i})=\frac{1}{2}\sum\limits _{j=1}^{z_{i}}(\mathbf{s}_{i}\cdot\mathbf{u}_{ij})^{2}$,
where $z_{i}$ is the coordination number at site $i$ and $\mathbf{u}_{ij}$
a unit vector connecting the nearest neighbors $i,j$. The constant
$K_{i}>0$ is usually denoted by $K_{{\rm c}}$ if the site $i$ is
in the core and by $K_{{\rm s}}$ if it is on the boundary.

However, it is a very difficult, if not impossible, task to deal with
the dynamics, especially the calculation of the relaxation rate and
magnetization reversal. Indeed, it is a formidable task to perform
a detailed analysis of the various critical points (minima, maxima,
saddle points) of the energy which is required for the study of the
relaxation processes.

In fact, it has been shown that, under some conditions which are quite
plausible for today's state-of-the-art grown nanomagnets, namely of
weak surface disorder, one can map this atomistic approach onto a
macroscopic model for the net magnetic moment
\begin{equation}
\bm{m}_{i}=\frac{1}{\mathcal{N}}\sum_{i=1}^{\mathcal{N}}\mathbf{s}_{i}\label{eq:Macrospin}
\end{equation}
of the NM evolving in the effective potential (H.O.T. = higher-order
terms)
\begin{equation}
\mathcal{E}_{{\rm eff}}=-K_{2}m_{z}^{2}+K_{4}\left(m_{x}^{4}+m_{y}^{4}+m_{z}^{4}\right)+\mbox{H.O.T}.\label{eq:Energy-EOSP}
\end{equation}

In the sequel we will refer to Eqs. (\ref{eq:Energy-EOSP}, \ref{eq:Macrospin})
as the effective-one-spin problem (EOSP).

The leading terms have coefficients $K_{2}$ and $K_{4}$ that are
functions of the atomistic parameters ($J,K_{{\rm c}},K_{{\rm s}},z,$
etc) and of the size and shape of the NM \citep{garkac03prl,kachkachi07j3m,Garanin_PhysRevB.98.054427}.
Note that both the core and surface may contribute to $K_{2}$ and
$K_{4}$. For example, when the anisotropy in the core in Eq. (\ref{eq:Ham-MSP})
is uniaxial, $K_{2}\simeq K_{{\rm c}}N_{{\rm c}}/\mathcal{N}$, where
$N_{{\rm c}}$ is the number of atoms in the core, see Ref. \citealp{kacbon06prb}.
In fact, even in this case the quartic term appears and is a pure
surface contribution. Regarding the coefficient of the quartic contribution
($k\equiv K/J$), for a sphere we have $k_{4}=\kappa k_{s}^{2}/zJ$
where $\kappa$ is a surface integral \citep{garkac03prl} and for
a cube $k_{4}=\left(1-0.7/\mathcal{N}^{1/3}\right)^{4}k_{s}^{2}/zJ$
\citep{Garanin_PhysRevB.98.054427}.

The most relevant parameter of this effective model is the ratio
\begin{equation}
\zeta\equiv K_{4}/K_{2}\label{eq:zeta_definition}
\end{equation}
which roughly represents the relative contribution of the surface
disorder and the ensuing spin noncolinearities. The details of the
conditions under which this model is applicable are discussed in Ref.
\citealp{garkac03prl} and may be summarized as follows. In order
to plot the energy of a many-spin NM we introduce a Lagrange parameter
which constrains the net magnetic moment to follow a specified path
in its phase space spanned by the spherical coordinates $\left(\theta,\varphi\right)$.
For this to be applicable the spin misalignment (or canting) should
not too strong. Therefore, the effective model can be built for NM
whose SA, as compared to the spin-spin exchange coupling, is not  too
strong. On the other hand, the NM size should not be too large for
it to be considered as a single magnetic domain, for a given underlying
material.

\begin{figure}
\includegraphics[angle=-90,scale=0.3]{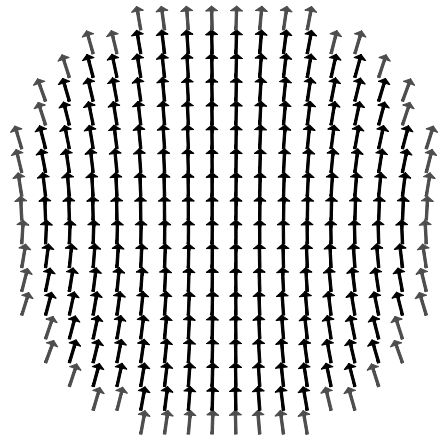}

\caption{Spin structure of linear dimension $N=20$.}
\end{figure}

To sum up, we have at hand a macroscopic model whose dynamics can
be studied using the full-fledged theory of Brown (or Langer) developed
during the last decades by many authors. This is nothing new. But
the new thing about this EOSP model is that it captures the main intrinsic
features of the NM through the coefficients of its effective potential
energy (\ref{eq:Energy-EOSP}). In particular, we can investigate
the effect of surface anisotropy on the relaxation rate of an individual
NM. This was done in Ref. \citealp{dejardinetal08jpd}. Indeed, in
Fig. 2 of this reference, the relaxation rate $\Gamma$ is plotted
against the parameter $\zeta$, as computed analytically using Langer's
(and Brown's) theory and numerically using the continued fraction
method. $\Gamma$ first increases as $\zeta$ increases, since in
this case the quartic term in Eq. (\ref{eq:Energy-EOSP}) creates
saddle points at the equator, but the minima are still determined
by the uniaxial anisotropy (the quadratic term). Beyond a critical
value of $\zeta$ the energy landscape becomes that of a cubic anisotropy,
namely the uniaxial-anisotropy minima become maxima and the new minima
are at the diagonals ($\pm1/\sqrt{3},\pm1/\sqrt{3},\pm1/\sqrt{3}$).
As $\zeta$ further increases these minima become deeper and the energy
barrier higher, thus leading to a decrease of $\Gamma$. A measurable
observable that can be built out of $\Gamma$ is the switching field
(SF), \emph{i.e.} the field at which the magnetization of the NM switches
at a given temperature and a given direction of the external magnetic
field. Upon varying the orientation of the latter, we obtain the so-called
Stoner-Wohlfarth astroid or the curve of the limit of metastability.
The measurement and discussion of the various aspects of the SF in
various situations of anisotropy can be found in the review \citealp{wernsdorfer01acp}.
Experimental observation of the magnetization reversal depends on
the relaxation time of the cluster and on the measuring time $\tau_{m}$
of the experimental setup. The magnetization reversal can be observed
only if the relaxation time is in the time window of the experiment,
or equivalently, if the relaxation rate is equal to the measuring
frequency $\nu_{m}=\tau_{m}^{\text{\textminus}1}$ . Therefore, for
the experimental observation of the magnetization reversal at finite
temperature, the relaxation rate as a function of the reduced anisotropy
barrier \footnote{The anisotropy constant $K$ used in $\sigma$ is the effective uniaxial
anisotropy constant.}
\begin{equation}
\sigma=KV/k_{{\rm B}}T,\label{eq:sigma_barrier}
\end{equation}
$\zeta$, the SF $h_{{\rm s}}$, the angle $\psi$ between the applied
field and the anisotropy axis and the damping parameter $\alpha$,
must be equal to the measuring frequency $\nu_{m}$, \emph{i.e.}

\begin{equation}
\Gamma(\sigma,\zeta,h_{{\rm s}},\psi,\alpha)=\nu_{m}.\label{eq:SF-condition}
\end{equation}

This equation can then be (numerically) solved for $h_{{\rm s}}$
in terms of $\nu_{m},\sigma,\zeta,\alpha$ and $\psi$. Obviously,
at very low temperature and $\zeta=0$ this yields the SW astroid
with the main feature that it homothetically shrinks as the temperature
increases and reaches reach zero at the blocking temperature. This
was confirmed by experiments on single cobalt NM using the microSQUID
technique \citep{jametetal01prl}.

For finite values of $\zeta$ one can also build the parametric plot
of the two components of the SF, namely $h_{x}=h_{{\rm s}}\sin\psi,h_{z}=h_{{\rm s}}\cos\psi$.
The result is given in Fig. 2 of Ref. \citealp{dejardinetal08jpd}.
The new feature that this plot shows is that the effect of finite
$\zeta$ (\emph{i.e.} that of surface anisotropy) is a flattening
of the astroid along one direction. For an experimental confirmation
of such a flattening of the SF curves, see the results in Fig. 6 of
Ref. \citealp{troetal03jmmm} obtained for $8\,{\rm nm}$ maghemite
NM. We note that this flattening of the SF curve owing to SA was observed
in Refs. \citealp{dimkac02jap,kacdim02prb}, where such a curve was
computed at very low temperature, using the model in Eq. (\ref{eq:Ham-MSP}),
for a spherical many-spin particle with local (on-site) anisotropy,
uniaxial in the core and transverse on the surface with a variable
constant.

\section{\label{sec:NMA-FS}Assembly of finite-size nanomagnets}

Here we briefly discuss another important extension that is relevant
to an interacting ANM of given size and shape. More precisely, we
would like to emphasize that, when dealing with DI between NM, even
in the macrospin approximation\emph{ }(ignoring SE), one has to take
account of their size and shape. Ignoring these parameters is what
is done in the so-called point-dipole approximation (PDA) which is
rather poor and may even lead to wrong results in situations with
NM too close to each other. Think of two magnetic layers that are
infinitely close to each other, for which the PDA predicts a finite
dipolar coupling energy while in reality the dipolar coupling between
the planes goes to zero, see Refs. \citealp{varalt00prb,francoetal14jap}.

In brief, for two finite-size elements, say of cylindrical symmetry
and of diameter $2R$ and thickness $2t$ with a center-to-center
distance $d$, one subdivides them into (differential) magnetic elements
$d\bm{m}_{i},i=1,2$, assumed to be point dipoles, and then writes
the energy of their DI as follows
\[
d\mathcal{E}_{{\rm DI}}=\left(\frac{\mu_{0}}{4\pi}\right)\frac{d\bm{m}_{1}\cdot d\bm{m}_{2}-3\left(d\bm{m}_{1}\cdot{\bf e}_{12}\right)\left(d\bm{m}_{2}\cdot{\bf e}_{12}\right)}{r_{12}^{3}}.
\]

For any two such NM within the assembly we write
\[
d\mathcal{E}_{{\rm DI}}^{\left(i,j\right)}=-\left(\frac{\mu_{0}}{4\pi}\right)d\bm{m}_{i}\cdot{\cal D}_{ij}d\bm{m}_{j}
\]
where ${\cal D}_{ij}$ is the DI tensor

\begin{equation}
{\cal D}_{ij}\equiv\frac{1}{r_{ij}^{3}}\left(3{\bf e}_{ij}{\bf e}_{ij}-1\right),\qquad{\bf r}_{ij}={\bf r}_{i}-{\bf r}_{j},\ {\bf e}_{ij}=\frac{{\bf r}_{ij}}{r_{ij}}.\label{eq:DDITensor}
\end{equation}

The next step consists in integrating over the whole volume of each
NM leading to the DI energy \citep{francoetal14jap}
\begin{equation}
\mathcal{E}_{{\rm DI}}=\eta\left[\bm{s}_{1}\cdot J_{12}\bm{s}_{2}-3\psi_{12}\left(\bm{s}_{1}\cdot\bm{e}_{12}\right)\left(\bm{s}_{2}\cdot\bm{e}_{12}\right)\right]\label{eq:Eint-Compactform}
\end{equation}
 where the (diagonal) matrix $J_{12}$ and the coefficient $\psi_{12}$
are given by
\begin{equation}
\left\{ \begin{array}{lll}
J_{12}=I, & \psi_{12}=1, & \mbox{vertical setup},\\
J_{12}=\left(\begin{array}{ccc}
1 & 0 & 0\\
0 & 1 & 0\\
0 & 0 & \Phi_{12}
\end{array}\right), & \psi_{12}=\frac{2+\Phi_{12}}{3}, & \mbox{horizontal setup},
\end{array}\right.\label{eq:AnisotropyDDICoeff}
\end{equation}
and $\lambda$ is the (dimensionless) coupling constant
\begin{equation}
\eta=\left(\frac{\mu_{0}}{4\pi}\right)\frac{\left(M_{s}V\right){{}^2}/d^{3}}{KV}\times\left\{ \begin{array}{lll}
\mathcal{I}_{s}^{\mathrm{v}}\left(\varrho\tau\right), &  & \mbox{vertical planes},\\
\\
\mathcal{I}_{d}^{\mathrm{v}}\left(\varrho,\tau\right), &  & \mbox{vertical disks},\\
\\
\mathcal{I}_{d}^{\mathrm{h}}\left(\varrho,\tau\right) &  & \mbox{horizontal disks}.
\end{array}\right.\label{eq:xi_over_sigma}
\end{equation}
and
\[
\varrho\equiv\frac{d}{2t},\quad\tau\equiv\frac{t}{R}.
\]

For later reference, we introduce the DI parameter $\lambda$
\begin{equation}
\lambda\equiv\left(\frac{\mu_{0}}{4\pi}\right)\frac{\left(M_{s}V\right)\text{\texttwosuperior}/d^{3}}{KV}.\label{eq:DI-param}
\end{equation}

Finally, $\Phi_{12}\left(\varrho,\tau\right)\equiv\mathcal{J}_{d}^{\mathrm{h}}\left(\varrho,\tau\right)/\mathcal{I}_{d}^{\mathrm{h}}\left(\varrho,\tau\right)$.
$\mathcal{I}_{d}^{\mathrm{h}}\left(\varrho,\tau\right)$ and $\mathcal{J}_{d}^{\mathrm{h}}\left(\varrho,\tau\right)$
are two shape integrals given explicitly by Beleggia \emph{et al.}
\citep{beleggiaetal04jmmm278}

Note that the dipolar energy (\ref{eq:Eint-Compactform}) depends,
on one hand, on the size and shape of the NM and on the other, on
their spatial arrangement via the parameters $\rho,\tau$, and $\xi$.
For example, the integral $\mathcal{I}_{s}^{\mathrm{v}}$ that obtains
for two vertically stacked planes, of lateral dimension $L$ and a
distance $d$ apart, reads ($\delta\equiv d/L$) \citep{varalt00prb,francoetal14jap}

\begin{align*}
\mathcal{I}_{s}^{\mathrm{v}}\left(\delta\right) & =4\delta^{3}\left[\begin{array}{c}
\delta-2\sqrt{1+\delta^{2}}+\sqrt{2+\delta^{2}}\\
\\
+\frac{1}{2}\log\left(1+\frac{1}{\delta^{2}}\right)+\log\left(\frac{1+\sqrt{1+\delta^{2}}}{1+\sqrt{2+\delta^{2}}}\right)
\end{array}\right].
\end{align*}

For small values of $\delta$ the integral $\mathcal{I}_{s}^{\mathrm{v}}\left(\delta\right)$
increases with $\delta$ as $4\delta^{3}\left[\sqrt{2}-2+\log\left(\frac{2}{\delta\left(1+\sqrt{2}\right)}\right)\right]$
while for large $\delta$ it does so as $1-\delta^{-2}+17\delta^{-4}/16$.
We thus see that as the distance between the two magnetic layers becomes
very small, \emph{i.e.} $\delta\rightarrow0$, the integral $\mathcal{I}_{s}^{\mathrm{v}}\left(\delta\right)$
and thereby the DI vanishes as it should. This result cannot be obtained
within the PDA for which $J_{12}$ and $\psi_{12}$ in Eq. (\ref{eq:Eint-Compactform})
are equal to unity. On the other hand, as the two magnets are very
far apart, \emph{i.e. }$\delta$ becomes very large, $\mathcal{I}_{s}^{\mathrm{v}}\left(\delta\right)\rightarrow1$
and thereby the DI reaches the limit of the PDA.

\section{\label{sec:SN-vs-NMA}Single NM versus assembly}

We have now presented the two ingredients that are necessary for studying
the competition between the SNM features and the collective behavior
in an array of NM. We have chosen to do so through three observables,
namely the equilibrium magnetization as a function of the external
magnetic field, the ac susceptibility and the FMR spectrum.

\subsection{Equilibrium magnetization}

Here we discuss some old experimental results regarding the behavior
of the magnetization of ANM as a function of the applied magnetic
field at varying temperature. This is our first example of compensation
between SE and DI.

Measurements of the magnetization at high fields performed on the
$\gamma$-Fe$_{2}$O$_{3}$ nanoparticles \citep{ezzir98phd,troncetal00jmmm}
and on cobalt particles \citep{cheetal95prb} showed that the magnetization
is strongly influenced by surface effects, as was evidenced by the
experimental study with variable particle size, see Fig. 3 of Ref.
\citep{troncetal00jmmm} and by (Monte Carlo) numerical investigation
of Ref. \citep{kacetal00epjb}. One finds that the $M(H)$ curves
at different temperatures present a rather different behavior as we
compare dilute with concentrated assemblies. Indeed, in Ref. \citealp{troncetal00jmmm}
(Fig. 3) the $M\left(H\right)$ curves look regular from $300\,K$
down to $100\,K$, split below $100\,K$ and instead of saturating,
the magnetization increases with increasing field and decreasing temperature.
The low-temperature increase starts at $100-75\,K$ and becomes steeper
as the temperature decreases. This steep increase is less important
with the increasing particle size, in accordance with the study of
SE in Ref. \citealp{kacetal00epjb}. A similar behavior is observed
with increasing concentration. For more concentrated dispersions the
low-temperature effect is still present but less marked, compared
to the corresponding dilute samples.

This interplay between surface anisotropy (an effect that is intrinsic
to the NM) and the DI (an effect pertaining to the assembly) was studied
in Refs. \citealp{margarisetal12prb,sabsabietal13prb}, using both
Monte Carlo simulations and analytical developments for dilute assemblies.

In Ref. \citealp{kacaze05epjb} this was performed for an assembly
of $\mathcal{N}$ ferromagnetic NM each carrying a magnetic moment
${\bf m}_{i}=m_{i}{\bf s}_{i},\,i=1,\cdots,{\cal N}$ of magnitude
$m_{i}$ and direction ${\bf s}_{i}$, with $\vert{\bf s}_{i}\vert=1$.
${\bf m}_{i}$ was defined in terms of the Bohr magneton $\mu_{B}$,
\emph{i.e.} $m_{i}=n_{i}\mu_{B}$, and $n_{i}$ are either all equal
for monodisperse assemblies or chosen according to some distribution.
Then, it was shown that in a dilute assembly, the magnetization of
a NM at site $i$ in a magnetic field applied in the $z$ direction,
to first order in the DI parameter, reads
\begin{equation}
\left\langle s_{i}^{z}\right\rangle \simeq\langle s_{i}^{z}\rangle_{0}+\sum_{k=1}^{\mathcal{N}}\xi_{ik}\langle s_{k}^{z}\rangle_{0}A_{ki}\frac{\partial\langle s_{i}^{z}\rangle_{0}}{\partial x_{i}},\label{eq:DDIMag}
\end{equation}
where $A_{kl}={\bf e}_{h}\cdot{\cal D}_{kl}\cdot{\bf e}_{h}$. In
Eq. (\ref{eq:DDIMag}) $\left\langle .\right\rangle $ is the statistical
average of the projection on the field direction of the particle's
magnetic moment. The following useful dimensionless physical parameters
are introduced

\begin{align*}
\xi_{ik}=\frac{\mu_{0}}{4\pi}\frac{m_{i}m_{k}/d^{3}}{k_{B}T},\quad x_{i} & \equiv\frac{n_{i}\mu_{B}H}{k_{B}T}=n_{i}x,\quad\sigma_{i}\equiv\frac{KV_{i}}{k_{B}T}.
\end{align*}

It is worth emphasizing that Eq. (\ref{eq:DDIMag}) yields the (local)
magnetization $\left\langle s_{i}^{z}\right\rangle $ of an interacting
assembly in terms of its ``free'' (with no DI) magnetization $\langle s_{i}^{z}\rangle_{0}$
and susceptibility $\partial\langle s_{i}^{z}\rangle_{0}/\partial x_{i}$,
with of course the contribution of the assembly ``superlattice''
via the lattice sum. As such, in order to investigate the competition
between DI and SE for example, one has to use for the SNM an expression
for the magnetization $\langle s_{i}^{z}\rangle_{0}$ that takes account
of SE. Restricting ourselves to monodisperse assemblies so as to investigate
the interplay between intrinsic and collective effects in pure form.
Then, $x_{i}=x,\sigma_{i}=\sigma,\xi_{ij}=\xi=\lambda\sigma$ {[}see
Eq. (\ref{eq:DI-param}){]} and Eq. (\ref{eq:DDIMag}) simplifies
into

\begin{equation}
\left\langle s^{z}\right\rangle \simeq m^{\left(0\right)}\left[1+\xi{\cal C}^{(0,0)}\frac{\partial m^{\left(0\right)}}{\partial x}\right].\label{eq:Mag_DDI}
\end{equation}
The lattice sum $\mathcal{C^{\mathrm{(0,0)}}}$ is the first of the
hierarchy of lattice sums (see Ref. \citep{kacaze05epjb}). For large
samples, we have ${\cal C}^{(0,0)}=-4\pi\left(D_{z}-1/3\right)$,
where $D_{z}$ is the demagnetizing factor in the $z$ direction.
It turns out that the relevant DI parameter, to this order of approximation,
is in fact 
\[
\tilde{\xi}\equiv\xi\mathcal{C}^{\left(0,0\right)}=\xi\times\frac{1}{\mathcal{N}}\sum_{i,j=1,i\neq j}^{{\cal N}}A_{ij}.
\]
The longitudinal susceptibility $\chi_{\parallel}^{\left(0\right)}=\partial m^{\left(0\right)}/\partial x$
is given in Ref. \citep{garpal00acp} and, injected in Eq. (\ref{eq:Mag_DDI})
leads to the approximate expression for the magnetization of a particle
taking account of its DI with the other particles in the assembly
\begin{equation}
\left\langle s^{z}\right\rangle \simeq m^{\left(0\right)}\left[1+\tilde{\xi}\left(a_{0}^{\left(1\right)}-\left(m^{\left(0\right)}\right)^{2}\right)\right].\label{eq:Mag_DDI_explicit}
\end{equation}
Where $a_{0}^{\left(1\right)}$ can be computed in the high-energy
barrier limit $a_{0}^{(1)}\left(\sigma\gg1\right)\simeq1-\frac{1}{\sigma}$.

Note that in the absence of any interactions or anisotropy, or at
high temperature, the magnetization is given by the Langevin function

\begin{equation}
\langle s^{z}\rangle_{0}\left(\sigma\rightarrow0,\xi\rightarrow0\right)=\mathcal{L}\left(\frac{\mu_{0}HM_{S}}{k_{B}T}\right).\label{eq:Langevin}
\end{equation}

Now, using the EOSP model (\ref{eq:Energy-EOSP}) for the individual
NM we compute the magnetization $m^{\left(0\right)}$ using perturbation
theory for small $\zeta=K_{4}/K_{2}$, see Ref. \citealp{sabsabietal13prb}
for the detailed derivation.

At first order in $\tilde{\xi}$, the equilibrium susceptibility reads

\selectlanguage{american}%
\begin{equation}
\chi^{\mathrm{eq}}\left(x,\sigma,\zeta,\tilde{\xi}\right)\simeq\chi_{\mathrm{free}}^{\mathrm{eq}}+\tilde{\xi}\chi_{\mathrm{int}}^{\mathrm{eq}}\label{eq:XiEq}
\end{equation}
where $\chi_{\mathrm{free}}^{\mathrm{eq}}$ is linear susceptibility
of the non-interacting assembly in the limit of high anisotropy energy
barrier \citep{sabsabietal13prb,jongar01prb}
\begin{eqnarray}
\chi_{\mathrm{free}}^{\mathrm{eq}}\left(x,\sigma,\zeta\right) & = & 2\chi_{0}^{\perp}\sigma\left[\chi_{\mathrm{free}}^{\left(1\right)}+3\chi_{\mathrm{free}}^{\left(3\right)}x^{2}\right],\label{eq:XiEqFree}\\
\nonumber \\
\chi_{\mathrm{free}}^{\left(1\right)} & = & \left(1-\frac{1}{\sigma}\right)+\frac{\zeta}{\sigma}\left(-1+\frac{2}{\sigma}\right),\nonumber \\
\chi_{\mathrm{free}}^{\left(3\right)} & = & \frac{1}{3}\left[\left(-1+\frac{2}{\sigma}\right)+\frac{\zeta}{\sigma}\left(2-\frac{5}{\sigma}\right)\right].\nonumber 
\end{eqnarray}
Here $\chi_{0}^{\perp}$ is the transverse equilibrium susceptibility
per spin at zero temperature in the absence of a bias field
\[
\chi_{0}^{\perp}\equiv\left(\frac{\mu_{0}m^{2}}{2KV}\right).
\]

The contribution of DI to the equilibrium susceptibility is given
by \citep{sabsabietal13prb}
\begin{eqnarray}
\chi_{\mathrm{int}}^{\mathrm{eq}}\left(x,\sigma,\zeta\right) & = & 2\chi_{0}^{\perp}\sigma\left[\chi_{\mathrm{int}}^{\left(1\right)}+3\chi_{\mathrm{int}}^{\left(3\right)}x^{2}\right],\label{eq:XiEqIntContr}\\
\nonumber \\
\chi_{\mathrm{int}}^{\left(1\right)} & = & 1-\frac{2}{\sigma}-2\left(1-\frac{3}{\sigma}\right)\frac{\zeta}{\sigma},\nonumber \\
\chi_{\mathrm{int}}^{\left(3\right)} & = & -\frac{4}{3}\left[\left(1-\frac{3}{\sigma}\right)-\frac{3\zeta}{\sigma}\right].\nonumber 
\end{eqnarray}

\selectlanguage{english}%
Therefore, the final result is the following asymptotic expression
for the magnetization taking account of both SA ($\zeta$) and DI
($\tilde{\xi}$), in addition of course to the contributions from
the uniaxial anisotropy ($\sigma$) and magnetic field ($x$)

\begin{align}
m\left(x,\sigma,\zeta,\tilde{\xi}\right) & \simeq\tilde{\chi}^{\left(1\right)}x+\tilde{\chi}^{\left(3\right)}x^{3}\label{eq:MagEOSPvsDDI1}
\end{align}
where

\begin{align}
\tilde{\chi}^{\left(1\right)} & \simeq\chi_{\mathrm{free}}^{\left(1\right)}+\tilde{\xi}\left[1-\frac{2}{\sigma}-2\left(1-\frac{3}{\sigma}\right)\frac{\zeta}{\sigma}\right],\label{eq:MagEOSPvsDDI2}\\
\nonumber \\
\tilde{\chi}^{\left(3\right)} & \simeq\chi_{\mathrm{free}}^{\left(3\right)}-\frac{4}{3}\tilde{\xi}\left[\left(1-\frac{3}{\sigma}\right)-\frac{3\zeta}{\sigma}\right]\nonumber 
\end{align}
are respectively the linear and cubic susceptibilities corrected by
DI.

The asymptotic expression (\ref{eq:MagEOSPvsDDI1}) shows how SA competes
with DI. Indeed, the sign of the surface contribution with intensity
$\zeta$ plays an important role in the magnetization curve. However,
since it appears coupled to the DI $\tilde{\xi}$ parameter, which
contains information regarding the sample's shape, it is the overall
sign of $\tilde{\xi}\zeta$ that determines whether there is a competition
between SE and DI or if the changes in magnetization induced by the
intrinsic and collective contributions have concomitant effects. We
know that for oblate samples ($\tilde{\xi}<0$) DI tend to suppress
the magnetization, whereas for prolate samples ($\tilde{\xi}>0$)
they enhance it. On the other hand, the effect of SA should be discussed
in relation with the quadratic contribution with coefficient $K_{2}$
in Eq. (\ref{eq:Energy-EOSP}). Indeed, for $\zeta>0$ the energy
minima of the quartic contribution are along the cube diagonals while
for $\zeta<0$ they are along the cube edges. Hence, the uniaxial
anisotropy with an easy axis along the $z$ direction, competes with
the cubic anisotropy when $\zeta>0$ whereas the two anisotropies
have a concomitant effect when $\zeta<0$. Consequently, we see from
Eqs. (\ref{eq:MagEOSPvsDDI1}, \ref{eq:MagEOSPvsDDI2}) that SE and
DI may have opposite or concomitant effects depending on their respective
signs. On this basis, and upon plotting the magnetization as a function
of the applied field, for different values of $\zeta$, for both prolate
and oblate assemblies, a plausible interpretation of the experimental
results of Ref. \citealp{troncetal00jmmm}, discussed above, was reached
in terms of a screening of the SA effects by DI, assuming that for
the studied $\gamma$-Fe$_{2}$O$_{3}$ samples $\zeta>0$ and $\tilde{\xi}<0$.

The expression in Eq. (\ref{eq:MagEOSPvsDDI1}) for the magnetization
of a (weakly) interacting assembly of NM, taking account of surface
anisotropy, may be applied to a variety of samples if they can be
produced with several concentrations and NM sizes and shapes. In particular,
arrays of platelet with varying aspect ratio and separation should
provide the necessary conditions for checking these theoretical predictions.

We may obtain the condition under which the SE and DI annihilate each
other, at least to first order in $x$ (applied field). Indeed, using
Eq. (\ref{eq:XiEqFree}) we rewrite Eq. (\ref{eq:MagEOSPvsDDI2})
as $\tilde{\chi}^{\left(1\right)}=\left(1-\frac{1}{\sigma}\right)+\delta\tilde{\chi}^{\left(1\right)}$,
where the $1^{{\rm st}}$ term is simply the linear susceptibility
in the absence of both SE and DI, and $\delta\tilde{\chi}^{\left(1\right)}$
the correction

\begin{align*}
\delta\tilde{\chi}^{\left(1\right)} & \equiv\frac{1}{\sigma}\left(-1+\frac{2}{\sigma}\right)\zeta+\left(1-\frac{2}{\sigma}\right)\tilde{\xi}-\frac{2}{\sigma}\left(1-\frac{3}{\sigma}\right)\zeta\tilde{\xi}.
\end{align*}
For typical samples one has $1/\sigma\ll1$ and thereby, at first
order the screening condition

\begin{equation}
\lambda_{{\rm s.c}}\simeq\frac{1}{{\cal C}^{(0,0)}}\times\frac{\zeta}{\sigma^{2}}.\label{eq:Mag-SC}
\end{equation}

Considering the fact that the parameter $\lambda$ (or $\xi$) is
positive, \textcolor{black}{Eq. (\ref{eq:Mag-SC}) implies that a
competition between SA and DI can only occur if $\zeta$ and ${\cal C}^{(0,0)}$
are of the same sign, at all temperatures. This is the case when the
ANM has a prolate shape, }\textcolor{black}{\emph{i.e.}}\textcolor{black}{{}
${\cal C}^{(0,0)}>0$, and the individual NM have a surface anisotropy
that leads (in the ESOP model) to a cubic anisotropy with $\zeta>0$.
Likewise, this competition may also set in for a $2D$ array of nano-elements
for which $\zeta<0$. Indeed, in this case the (effective) cubic anisotropy
has minima along the normal to the array plane whereas the DI drive
the magnetization into the plane (since for such an oblate sample
${\cal C}^{(0,0)}<0$). Note also that the condition (\ref{eq:Mag-SC})
involves the temperature ($\lambda_{{\rm s.c}}\propto T^{2}$) and
this to be expected because it is derived from the magnetization which
is temperature dependent. In fact, this screening condition shows
that for a set of parameters, namely the underlying material ($K$),
the assembly spatial arrangement (${\cal C}^{(0,0)}$), the NM characteristics
($V,\zeta$), there is a compensation between DI and SE due to the
underlying processes related with spin fluctuations which are essentially
temperature-dependent.}

\subsection{AC susceptibility}

For the AC susceptibility we need to compute the relaxation rate of
the individual NM including both SE and DI, before we can study the
competition between the intrinsic features and collective behavior.
This was done for monodisperse assemblies in Ref. \citealp{vernayetal14prb}.

For a situation with a longitudinal field, \emph{i.e.} an assembly
with anisotropy axes oriented along the field ($\psi=0$), the transverse
response can be considered instantaneous ($\tau_{\perp}=0$). Hence,
with $\tau_{\parallel}=\Gamma^{-1}$ and the equilibrium susceptibility
$\chi_{\text{\ensuremath{\parallel}}}=\chi^{{\rm eq}}$ given in Eq.
(\ref{eq:XiEq}), the AC susceptibility reads (in the simplest model
of Debye)

\selectlanguage{american}%
\begin{equation}
\chi\left(x,\sigma,\zeta,\tilde{\xi},\eta\right)=\frac{\chi_{\parallel}^{\mathrm{eq}}}{1+i\omega\Gamma^{-1}}.\label{eq:XiACAssembly}
\end{equation}

\selectlanguage{english}%
The relaxation rate $\Gamma$ for a NM with surface effects and interacting
within the assembly, was computed in Ref. \citealp{vernayetal14prb}
and will not be reproduced here. Here we only discuss the results
obtained there regarding the competition under study. For example,
\foreignlanguage{american}{the results in Fig. 5 show that the surface
anisotropy, in the case of positive $\zeta$, has the opposite effect
compared to DI. This implies that surface effects can screen out the
effect of DI or the other way round. This confirms the results of
Ref. \citealp{sabsabietal13prb}, discussed in the previous section,
for equilibrium properties for both negative and positive $\zeta$.
We further refer the reader to Ref. \citealp{sabsabietal13prb} where
a detailed discussion of the competition between DI and SE was presented
in what regards the frequency dependence of the real and imaginary
parts of the AC susceptibility and the shift of their respective peaks.}

The screening condition here is rather difficult to establish in an
analytical form because the relaxation rate, which depends on both
$\zeta$ and $\xi$, is only known semi-analytically. Nevertheless,
we can establish a screening criterion with regards to the effective
temperature $\theta_{{\rm VF}}$ that is very often used in the Vogel-Fulcher
law to account for DI. Accordingly, in Ref. \citealp{vernayetal14prb}
we obtained the following analytical expression \foreignlanguage{american}{for
$\theta_{\mathrm{VF}}$
\begin{equation}
\frac{\theta_{\mathrm{VF}}}{T}=\frac{\zeta}{4}+\frac{1}{6\sigma}\left(\xi^{2}\mathcal{S}\right)\label{eq:ThetaVFCorrected}
\end{equation}
}where $\mathcal{S}$ is a function of the superlattice. Obviously,
in the absence of DI, and according to our formalism, in the absence
of SE as well, $\theta_{{\rm VF}}$ vanishes. However, according to
Eq. (\ref{eq:ThetaVFCorrected}), $\theta_{{\rm VF}}$ may vanish
even in the presence of both effects if $\zeta<0$. Indeed, in this
case we obtain the screening condition ($\zeta=-\left|\zeta\right|$)
\begin{equation}
\lambda_{{\rm s.c}}\simeq\sqrt{\frac{3}{2\mathcal{S}}}\times\left(\frac{\left|\zeta\right|}{\sigma}\right)^{1/2}.\label{eq:TVF-SC}
\end{equation}

Similarly to the condition (\ref{eq:Mag-SC}), derived from the equilibrium
magnetization, the present condition involves the assembly superlattice
($\mathcal{S}$) and the NM parameters ($K,V,\zeta$) and temperature
with a different power. 

\subsection{Ferromagnetic resonance}

Now we come to the last example we would like to discuss ; this is
the ferromagnetic resonance (FMR) of an interacting ANM. In particular,
we investigate the shift of the corresponding frequency induced by
SE and DI. In order to derive an approximate expression for the frequency
shift due to DI, we consider an array with large enough inter-NM separation
so that we can make use of perturbation theory. Then, we split the
total energy into a free part plus the interaction contribution \citep{DejardinEtal_PhysRevB.97.224407}
\[
\mathbb{H}=\mathcal{F}+\mathcal{I}=\mathcal{F}\left(1+\mathcal{F}^{-1}\mathcal{I}\right).
\]

The product $\mathcal{F}^{-1}\mathcal{I}$ scales with the ratio $\lambda/H$,
\emph{i.e. }the ratio of the DI intensity to the static magnetic field
$H$. This ratio is obviously small for a dilute assembly, especially
for standard FMR measurements where the DC field is usually taken
strong enough to saturate the sample (usually between $0.3\,\mathrm{T}$
and $1\,{\rm T}$).

\subsubsection{Without surface effects}

For a $2D$ array we obtain the explicit DI correction to the FMR
(dimensionless) angular frequency $\varpi$ of the array of NM 
\begin{equation}
\Delta\varpi_{{\rm {\rm DI}}}\simeq-\frac{\lambda}{2\kappa^{3}}\times\frac{1}{\mathcal{N}}\sum_{i=1}^{\mathcal{N}}\overset{\mathcal{N}}{\sum_{j=1,j\neq i}}\,\frac{\mathcal{J}_{d}^{\mathrm{h}}\left(\eta_{ij},\tau\right)}{r_{ij}^{3}}\label{eq:DICorrection-v2}
\end{equation}
where $\varpi\equiv\omega/\omega_{K}$, with $\omega_{K}\equiv\gamma H_{K}$,
and $\mu_{0}H_{K}=2K_{2}/M_{s}$ the anisotropy field. In addition,
we introduce the parameters

\[
\kappa\equiv\frac{d}{2R},\quad\eta_{ij}=\left(\frac{d}{L}\right)r_{ij}.
\]

\begin{figure}
\begin{centering}
\includegraphics[width=0.95\columnwidth]{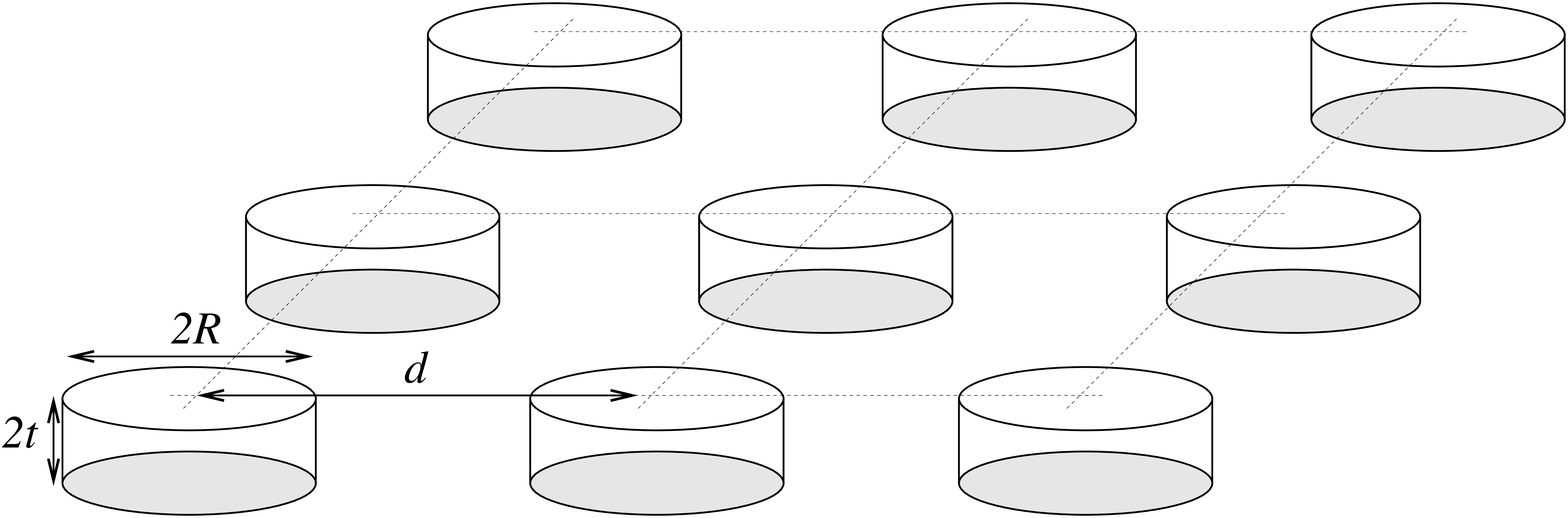}
\par\end{centering}
\caption{\label{fig:Platelets}Lattice of thin disks with lattice parameter
$d$.}

\end{figure}

For an array of platelets with $\tau=t/R\ll1$, as in Fig. \ref{fig:Platelets},
the expression above simplifies into

\begin{equation}
\Delta\varpi_{{\rm {\rm DI}}}\simeq\Delta\varpi_{{\rm pda}}-\frac{9}{32}\frac{\lambda}{\kappa^{5}}\mathcal{C}_{5}=\Delta\varpi_{{\rm pda}}\left(1+\frac{9}{16\kappa^{2}}\frac{\mathcal{C}_{5}}{\mathcal{C}_{3}}\right)\label{eq:DICorrection-Disks-2}
\end{equation}
where we have singled out the contribution $\Delta\varpi_{{\rm pda}}\equiv-\frac{\lambda}{2\kappa^{3}}\mathcal{C}_{3}$
that obtains within the PDA \footnote{Note that $\mathcal{C}_{3}$ vanishes for cubes and spheres and for
such systems only the first expression in Eq. (\ref{eq:DICorrection-Disks-2})
should be used. }, with the usual lattice sums
\[
\mathcal{C}_{n}\equiv\frac{1}{\mathcal{N}}\sum_{i=1}^{\mathcal{N}}\overset{\mathcal{N}}{\sum_{j=1,j\neq i}}\frac{1}{r_{ij}^{n}}.
\]

Several observations may be made. First, we see that while the PDA
shift of the resonance frequency behaves, as expected, namely as $1/d^{3}$,
the shift due to finite-size scales as $1/d^{5}$ for the array of
platelets. This contribution is obviously smaller than that of PDA,
but it increases as the NM come closer to each other.

\begin{figure}
\begin{centering}
\includegraphics[scale=0.27]{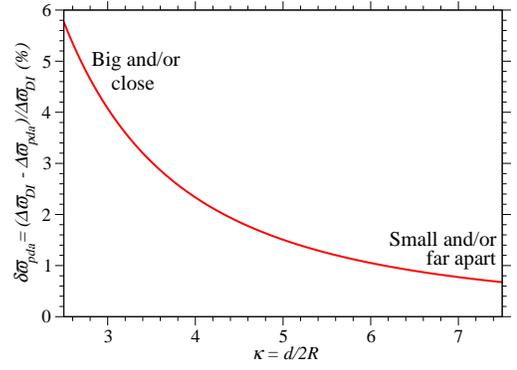}
\par\end{centering}
\caption{\label{fig:PDA-vs-FSDI}Relative difference between the PDA contribution
and that from the finite-size-DI to the FMR frequency.}
\end{figure}

Second, as can be seen in Fig. \ref{fig:PDA-vs-FSDI}, the PDA should
be applied only for small and/or well separated NM. To give an order
of magnitude we consider the FeV nanodisks studied in Refs. \citealp{mitsuzukaetal12apl,pigeauetal12prl}
with $2t=27\,{\rm nm},2R=600\,{\rm nm},d=1600\,{\rm nm}$.\textcolor{black}{{}
The results confirm that, for not-too-dense assemblies, }\textcolor{black}{\emph{i.e.}}\textcolor{black}{{}
for $2.5\lesssim\kappa\lesssim3$, there is a variation ($\delta\varpi_{{\rm pda}}\simeq5\%$)
of the frequency shift due to the fact that the NM are not simple
point dipoles. This variation should be accessible to experiments.
Obviously, for very dilute assemblies ($\kappa\gtrsim7$) the PDA
provides a quite reasonable description of the physics up to an error
less than 1\%.}

\subsubsection{Including surface effects}

As was shown in Ref. \citealp{DejardinEtal_PhysRevB.97.224407}, adding
the surface contribution to the free-particle energy according to
the effective model (\ref{eq:Energy-EOSP}) adds the term $-\zeta\sum_{\alpha=x,y,z}m_{i,\alpha}^{3}\bm{e}_{\alpha}$
to the effective field and thereby the total frequency shift, due
to both DI and SE, is given by
\begin{equation}
\Delta\varpi=-\varpi_{{\rm SE}}+\Delta\varpi_{{\rm {\rm DI}}}=\zeta-\frac{\lambda}{2\kappa^{3}}\left[\mathcal{C}_{3}+\frac{9}{16\kappa^{2}}\mathcal{C}_{5}\right].\label{eq:TotalFreqShift}
\end{equation}

Note that due to the SA contribution, the FMR frequency of a single
NM may either increase or decrease according to the sign of $\zeta$.
Consequently, for $\zeta>0$ we see that SA may compete with DI.

\begin{figure}
\begin{centering}
\includegraphics[width=0.95\columnwidth]{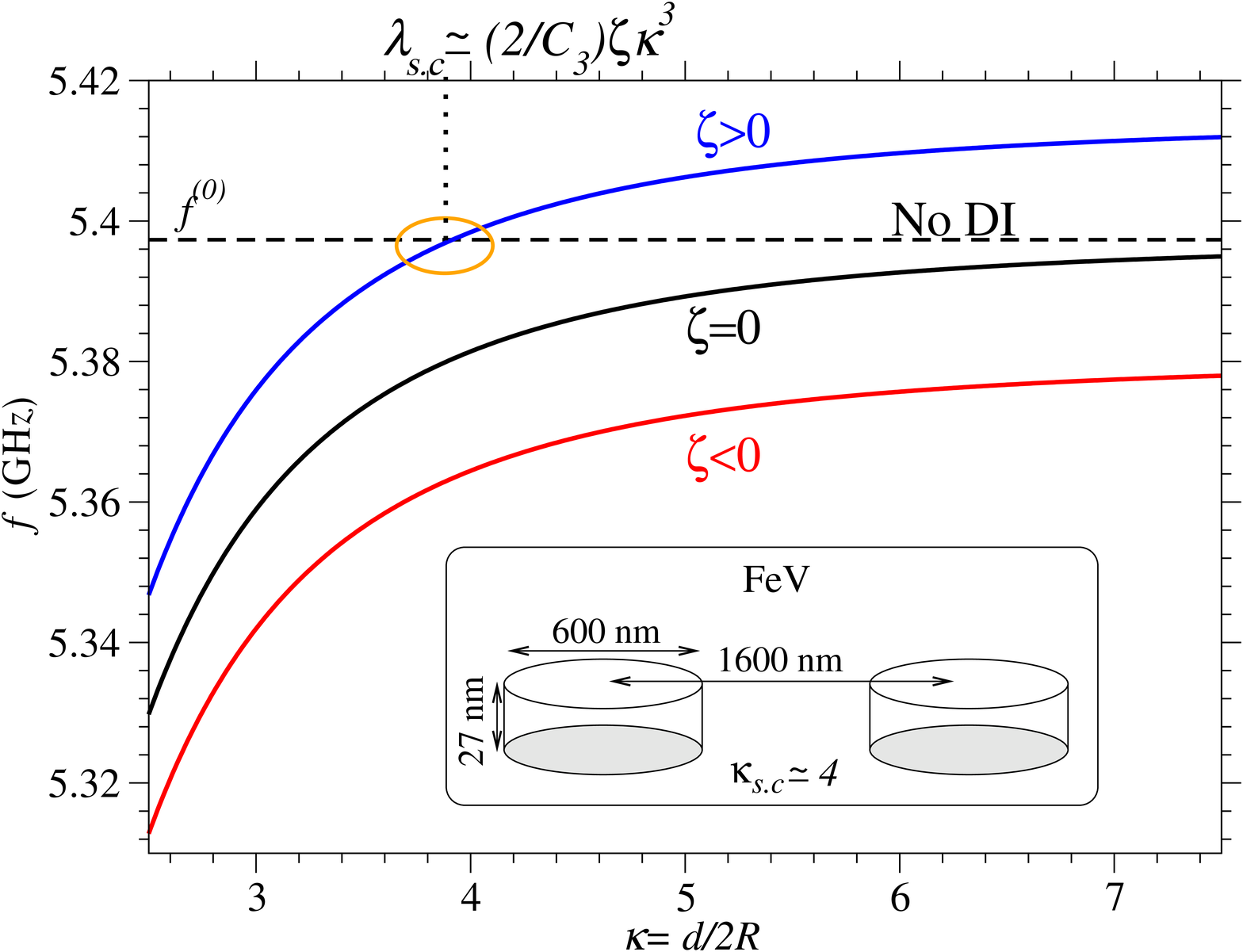}
\par\end{centering}
\caption{\label{fig:FMR_DI-vs-SE}FMR frequency as a function of the NM separation
for the noninteracting assembly (dashed line) and for interacting
assemblies with $\zeta>0$ (blue curve), $\zeta=0$ (black curve),
$\zeta<0$ (red curve).}
\end{figure}

Indeed, in Fig. \ref{fig:FMR_DI-vs-SE} we see that as the NM come
closer to each other (smaller $\kappa$) the FMR frequency of the
interacting array is red-shifted. On the opposite, when $\zeta$ becomes
positive (going from the black to the blue curve) the FMR frequency
is blue-shifted. Second, we see that at some value of $\zeta$ and
$\kappa$, \emph{i.e. }for some parameters of the NM on one hand,
and some parameters for the assembly, on the other, the FMR frequency
of the interacting array (blue curve) crosses that of the noninteracting
assembly (horizontal dashed line). The screening condition in this
case is given by
\begin{equation}
\lambda_{{\rm s.c}}\simeq\frac{2}{\mathcal{C}_{3}}\times\zeta\kappa^{3}.\label{eq:FMR-SC}
\end{equation}
\textcolor{black}{Contrary to the conditions (\ref{eq:Mag-SC}) and
(\ref{eq:TVF-SC}), the condition above only depends on the samples
properties and not on temperature. This is not surprising knowing
that the resonance frequency is an intrinsic property of the sample.} 

For the FeV disks considered above, we may invert the expression in
Eq. (\ref{eq:FMR-SC}) to find the numerical condition on $\kappa$;
we find $\kappa_{{\rm s.c.}}\simeq4$. Such a compensation point could
be achieved for some NM assemblies with reasonable physical parameters.
For instance, in Ref. \citealp{ancinasetal01jap}\citealp{ancinasetal01jap}
it was found that the resonance frequencies of dense assemblies of
NiFe nanowires could be recovered using Kittel's expression for noninteracting
assemblies {[}see Fig. 4 of the cited reference{]}. This means that
the DI of the dense assembly are compensated for such nanowires.

\section{Conclusions}

After summarizing the theoretical developments that allow us to take
into account both the intrinsic and collective effects in an assembly
of nanmagnets, we study \textcolor{black}{the competition between
surface anisotropy and dipolar interactions through three different
observables: the magnetization, the ac susceptibility, and FMR. For
each of these observables, we have demonstrated that there exists
a set of physical parameters, pertaining to the nano-elements themselves
and to their arrangement in an assembly, for which a compensation
between surface effects and inter-element interaction occurs. This
implies that the corresponding assembly would behave as noninteracting.
In each case, we have given an explicit expression for the screening
condition in terms of the physical parameters. The screening conditions
in Eqs. (\ref{eq:Mag-SC}) and (\ref{eq:TVF-SC}) differ significantly
from Eq. (\ref{eq:FMR-SC}) in the sense that the first two depend
on the temperature while the latter does not. This implies that for
a specifically synthesized sample the FMR frequency is a more suitable
observable for achieving noninteracting assemblies of dressed particles
over a large range of temperature. As for the magnetization and ac
susceptibility measurements, the temperature scaling laws for the
screening condition ($\lambda_{{\rm s.c}}\propto T^{2}$ or $\lambda_{{\rm s.c}}\propto T^{1/2}$
) imply that one does not need to specifically design a sample in
order to have a DI-SA compensation: this situation can be achieved
by simply estimating the temperature at which this situation should
occur.}

\section*{Glossary}

\noindent NM = nanomagnet(s), SNM = single nanomagnet(s), ANM = nanomagnet
assembly, SE = surface effects, SA = surface anisotropy, EOSP = effective-one-spin
problem, DI = dipolar interaction(s), PDA = point-dipole approximation.

\bibliography{hkbib}

\end{document}